\documentclass[12pt]{article}
\usepackage{geometry}  
\usepackage{graphicx}
\usepackage{amssymb}
\usepackage{amsmath}
\usepackage{epstopdf}
\usepackage{comment}

\usepackage[hyperindex=true,
          pdfstartview=FitH,
          bookmarksnumbered=true,
          bookmarksopen=true,
          citecolor=blue,
          linkcolor=blue,
          colorlinks=true,
          unicode]{hyperref}

\allowdisplaybreaks

\begin{document}

\title{Novel rotating hairy black hole in (2+1)-dimensions}
\author{Liu Zhao\thanks{Correspondence author}, Wei Xu and Bin Zhu\\
School of Physics, Nankai University, 
Tianjin 300071, China\\
{\em email}: \href{mailto:lzhao@nankai.edu.cn}{lzhao@nankai.edu.cn}, 
\href{mailto:xuweifuture@mail.nankai.edu.cn}{xuweifuture@mail.nankai.edu.cn} \\
and \href{mailto:binzhu7@gmail.com}{binzhu7@gmail.com}}
\date{}
\maketitle

\begin{abstract}

We present some novel rotating hairy black hole metric in $(2+1)$ dimensions, 
which is an exact solution to the field equations of the Einstein-scalar-AdS 
theory with a non-minimal coupling. The scalar potential is determined by the 
metric ansatz and consistency of the field equations and cannot be prescribed 
arbitrarily. In the simplified, critical case, the scalar potential contains two 
independent constant parameters, which are respectively related to the mass and 
angular momentum of the black hole in a particular way. As long as the angular 
momentum does not vanish, the metric can have zero, one or two horizons. The 
case with no horizon is physically uninteresting because of the curvature 
singularity lying at the origin. We identified the necessary conditions
for at least one horizon to be present in the solution, which imposes some bound 
on the mass-angular momentum ratio. For some particular choice of parameters our 
solution degenerates into some previously known black hole solutions. 
\vspace{3mm}

\vspace{3mm}

\end{abstract}

\section{Introduction} 

In spite of the fact that our spacetime is $(3+1)$-dimensional, gravity in $(2+1)$ 
spacetime dimensions is still an active field of study because of its relative
simplicity and its applications in realizing and testing some fundamental 
ideas which may apply to all spacetime dimensions. For instance, $(2+1)$-dimensional 
gravity is often invoked in the studies of AdS/CFT duality 
\cite{Henneaux:2002p2538,Hasanpour:2011p2274} 
and/or of basic properties black holes.
The first known and the most famous black hole solution in $(2+1)$ dimensions
is the Ba\~nados-Teitelboim-Zanelli (BTZ) black hole \cite{Banados:1992wn}, 
which is an exact solution of the 
$(2+1)$-dimensional Einstein equation in the absence of matter source. 
Charged rotating variants of BTZ black hole 
was found in \cite{Clement1, Clement2}, see also
\cite{Martinez:1999p2523}. It is often tempting and challenging to
include extra matter sources other than simply the Maxwell field and see whether 
novel black hole solutions will emerge and whether the nice properties
of the BTZ black hole and its charged rotating variants 
(such as the duality relationships) still persist. 
In this respect, an extra scalar source is the simplest choice and has been 
studied extensively.

The study of hairy extensions of the BTZ black hole has a history which is 
almost as long as the age of BTZ black hole itself. To this date, many explicit 
hairy extensions of BTZ black hole has been found, and in most cases the extra 
hair is a scalar field which couples to gravity either minimally 
\cite{Martinez:2004p2545,Correa:2010p2540,Correa:2011p2543,
Correa:2012p2547,Martinez:2006p2553,Singleton} or non-minimally 
\cite{Martinez:1996gn, Hortacsu:2003we}, the scalar may or
may not couple to itself via a scalar potential, and the hairy black holes can 
be static, charged or rotating, see \cite{Charged hairy} for more references on 
related studies. Charged static solutions are relatively 
easy to find, however it is very difficult to find charged rotating solutions 
due to the complexities involved in solving the Maxwell field equation in 
rotating background.

In this paper, we present some novel exact rotating black hole
solutions in $(2+1)$-dimensional gravity with a non-minimally coupled
scalar field. Rotating solution in a similar theory was previously studied in 
\cite{Hortacsu:2003we} in the absence of scalar potential. The metric ansatz 
used in \cite{Hortacsu:2003we} is different from ours, and hence the solution 
presented there is not the limit of our solution in the limit of vanishing 
scalar potential. In our case, the metric ansatz
together with consistency of the field equations determines the 
scalar potential up to five adjustable integration constants, one of which
is a negative cosmological constant as is required by the existence of a smooth 
event horizon \cite{Ida}. In a simplified branch of solutions, 
the scalar potential behaves like $\phi^6+\mathcal{O}(\phi^{10})$ when $\phi$ is 
small, and if the coefficient of the
$\phi^6$ term is negative, the true vacuum energy gets an extra negative 
contribution and will differ from the bare cosmological constant.

The paper is organized as follows. In Section 2, we describe the action of the 
model and the corresponding field equations. That the metric ansatz
together with consistency of Einstein's equation determine the 
scalar field and its potential some integration constants is shown clearly.
Fixing one of the constants to a special value, we are led to a much simplified,
critical branch of solution which is described in Section 3. This branch of 
solution contains an essential singularity at the origin 
which has to be enclosed by some event horizons and the whole spacetime is not 
conformally flat. The behavior of the scalar potential is also 
briefly discussed in this section.
In Section 4 we discuss the horizon structure and the conditions required to 
ensure the existence of horizons. If some particular values for the coupling 
strength for the scalar field or the integration constants in the metric are 
chosen, our solution 
will degenerate into some known solutions found in the literature. Meanwhile, 
some other specific choices of parameters may greatly simplify the solution and 
will allow us to discuss the properties of the solution in greater depth. These 
are discussed in Section 5. Finally, in Section 6, some concluding remarks will 
be given.

\section{Action, field equations and generic solution}

The model we consider is the $(2+1)$-dimensional gravity with a  
non-minimally coupled scalar field, which is described by the action
(omitting some appropriate boundary terms which are necessary to make the variation of the action consistent)
\begin{align} 
I=\frac{1}{2}\int\mathrm{d} x^3\sqrt{-g}\left[
  R -g^{\mu\nu}\nabla_{\mu}\phi\nabla_{\nu}\phi
  -\frac{1}{8} R\phi^2-2\,V(\phi)\right], \label{action} 
\end{align}
where we have set the gravitational constant $\kappa=8\pi G=1$, and the
coupling constant between gravity and the scalar field is fixed to be
$\xi=\frac{1}{8}$. The reason for this choice is that when the scalar
potential $V(\phi)$ is vanishing, the scalar action (i.e. the second and the 
third terms together) is conformally invariant.
On the other hand, when $V(\phi)$ is nonconstant, this particular
choice of gravity-scalar coupling allows us to obtain exact solutions
in a neat form.  

We did not prescribe a concrete form for the scalar potential $V(\phi)$.
The reason for this lies in that, unlike the cases in Minkowski spacetime, 
the scalar potential in curved spacetime cannot be prescribed arbitrarily, just 
like we cannot simply prescribe an arbitrary metric and announce that it is a 
solution to Einstein's equation by identifying the value of the corresponding
Einstein tensor as the stress-energy tensor of some unknown matter source. 
In real physical interesting situations, the stress-energy tensor of the 
matter source must have a good field theoretic or fluid mechanical description.
Likewise, a physically good matter source must allow for the existence of 
a solution for the metric field with reasonable symmetries.

In this paper, we are interested in studying solutions with a rotation symmetry.
To reach our goal, we first need to write down the field equations associated 
with the action. These are given by
\begin{align}
  &E_{\mu\nu}\equiv G_{\mu\nu}
  -T^{[\phi]}_{\mu\nu}+V(\phi)g_{\mu\nu}=0,  \label{EinEq}\\
  &\square\phi-\frac{1}{8}\, R\phi-\partial_{\phi}V(\phi)=0,
\end{align}
where
\begin{align*}
  T^{[\phi]}_{\mu\nu}&=\partial_{\mu}\phi\partial_{\nu}\phi
  -\frac{1}{2}g_{\mu\nu}\nabla^{\rho}\phi\nabla_{\rho}\phi
  +\frac{1}{8}\big(g_{\mu\nu}\square-\nabla_{\mu}\nabla_{\nu}+G_{\mu\nu}
  \big)\phi^2
\end{align*}
is the stress-energy tensor associated with the scalar field $\phi$. 

To obtain the rotating solution, we begin with the following metric ansatz
\begin{align}
  \mathrm{d} s^2=-f(r)\mathrm{d} t^2+\frac{1}{f(r)}
  \mathrm{d} r^2+r^2\bigg(\mathrm{d} \psi+\omega(r)\mathrm{d} t\bigg)^2,
  \label{lineelm}
\end{align}
where the coordinate ranges are given by $-\infty<t<\infty$, $r\geq0$,
$-\pi\leq\psi\leq\pi$. Using this metric ansatz, it is straightforward to get
\begin{align*}
16r\left(E_t{}^t-E_r{}^r\right) &=
r \big( -2{r}^{2}\,\phi
\phi_{,\,r}\, \omega \omega_{,\,r} +8 {r}^{2}\,\omega \omega_{,\,rr} +24r\, 
\omega \omega_{,\,r}+12f \,(\phi_{,\,r})^2 \\
& \quad -{r}^{2}\, \phi^{2}\, \omega\omega_{,\,rr} 
-3{r} \,\phi^{2}\, \omega\omega_{,\,r} 
-4f \,\phi \phi_{,\,rr} \big)=0.
\end{align*}
Meanwhile, we also have
\begin{align}
16E_\psi{}^t &=-r \big( -24\omega_{,\,r} -8r \, \omega_{,\,rr}
+2r\,\phi \phi_{,\,r} \, \omega_{,\,r} 
+3\phi^2\,\omega_{,\,r} + r\,\phi^2\, \omega_{,\,rr} \big)=0. 
\label{omegaeq}
\end{align}
These two equations together give
\begin{align*}
\phi\phi_{,\,rr}-3 (\phi_{,\,r})^2 =0,
\end{align*}
which yields the solution
\begin{align}
\phi(r)=\pm\frac{1}{\sqrt{kr+b}}, \label{scsol}
\end{align}
with $k$ and $b$ being integration constants. It is remarkable that the value of 
the scalar field is solely determined by the metric ansatz and the consistency 
of Einstein's equation, without referring to the scalar potential and the scalar 
field equation. Moreover, the dependance on the metric ansatz is quite weak, 
because the very same solution (\ref{scsol}) has also appeared in our previous 
work \cite{Charged hairy} on a charged static circularly symmetric solution.

Now inserting (\ref{scsol}) into (\ref{omegaeq}) and solve the resulting 
equation, we get
\begin{align}
\omega(r) &= c_1\bigg(\frac{ 8 k^2}{(8 b-1)^3}\log \frac{r}{8 b-1+8 k r}
-\frac{8 b^2-b-2 k r}{2 (8 b-1)^2 r^2}
\bigg)+c_2, \label{omega}
\end{align}
where $c_1$ and $c_2$ are integration constants. In order for the metric to
contain nontrivial rotation, we need $c_1 \ne 0$. The asymptotic behavior of 
$\omega(r)$ as $r\to+\infty$ reads
\begin{align*}
\omega(r) =c_2 -\frac{8k^2 \log(8k)}{(8b-1)^3}c_1 -\frac{c_1}{16} r^{-2}
-\frac{c_1}{192k} r^{-3}
+\mathcal{O}(r^{-4}).
\end{align*}
Thus, removal of global 
rotations of the coordinate system, i.e. 
$\omega(r)|_{r\to+\infty} =0$, lead us to the choice 
\begin{align}
c_2 =\frac{8k^2 \log(8k)}{(8b-1)^3}c_1 \label{c2}
\end{align}
provided $k\ne 0$. If $k=0$, then the scalar field $\phi$ is constant, and 
hence the action becomes that of the standard Einstein-Hilbert action with a 
cosmological constant. In this paper, we are not interested in this degenerated 
case.

Now consider the case $k\ne 0$. Inserting (\ref{c2}) into (\ref{omega})
and setting $c_1=8a$, we have
\begin{align}
\omega(r) &= 8a\bigg(\frac{ 8 k^2}{(8 b-1)^3}\log \frac{8kr}{8 b-1+8 k r}
-\frac{8 b^2-b-2 k r}{2 (8 b-1)^2 r^2}
\bigg). \label{omega1}
\end{align}
Therefore, we are left with only three independent constant parameters $a, b, k$ 
in $\omega(r)$. Here $a$ is a rotation parameter and is related to the 
angular momentum of the solution.
Inserting (\ref{scsol}), (\ref{omega1}) into the rest of the 
field equations, we can get a very complicated, yet exact solution for the 
metric function $f(r)$, together with a more complicated, exact scalar 
potential $V(\phi)$ which takes pages to be written down. 
These results are so complicated that we do not think 
it worth to reproduce the concrete expressions here. However, there are two 
important points to be noted:
\begin{itemize}
\item In the process of solving $f(r)$, two new independent integration
constants will arise, one of which also appear in the final scalar potential as 
a constant term. This constant can be easily identified to be related to 
the cosmological constant, and the existence of smooth black hole horizons
will force the cosmological constant to be negative and best denoted as 
$\Lambda=-\frac{1}{\ell^2}$,
where $\ell$ has the interpretation as AdS radius. The other integration 
constant arising in the solution for $f(r)$ may be denoted $\beta$ and is 
related to the mass parameter;
\item The form of the scalar potential $V(\phi)$ is completely determined by the 
metric ansatz and consistency of the field equations. The only adjustable part
of the potential is the set of integration constants that are inherent from the
scalar field and the metric functions $f(r)$ and $\omega(r)$. There are five 
such parameters, i.e. $a, b, k, \beta$ and $\ell$.
\end{itemize}
Due to the very complicated form of the solution, we are unable to proceed to 
make quantitative analysis on the solution for generic choices of the 
constants $a, b, k, \beta$ and $\ell$. 

Notice, however, that the solution (\ref{omega1}) is apparently divergent at 
$b = \frac{1}{8}$. From either the action or the field equations, 
there is no sign of divergence for this particular choice of integration 
constant. Since the parameter $b$ is related to the value of $\phi(r)$ at the 
origin, the apparent divergence at $b=\frac{1}{8}$ might signify 
some critical behavior of the solution. In order to understand more on this
point, let us look at the asymptotic behavior of $\omega(r)$ as 
$b\to \frac{1}{8}$. We have
\begin{align}
\omega(r) &=\frac{a k^2 \log \left(\frac{1}{8 k}\right)+a k^2 \log (8 k)}{8
   \left(b-\frac{1}{8}\right)^3}-\frac{a (12 k r+1)}{24 \left(k
   r^3\right)}+\frac{a\left(b-\frac{1}{8}\right) }{32 k^2 r^4}
   +\mathcal{O}\left(\left(b-\frac{1}{8}\right)^2\right). \label{omegab}
\end{align}
The first term on the right hand side actually vanishes because it is 
proportional to $\log(1)$ (provided $k\ne 0$). The third and rest terms 
are polynomial in $b-\frac{1}{8}$ and they all vanish at $b=\frac{1}{8}$.
Therefore, we are left with only the second term which is completely regular at 
$b=\frac{1}{8}$. One can check that the second term alone indeed solves 
eq.(\ref{omegaeq}) if the parameter $b$ is taken to be equal to $\frac{1}{8}$. 
Proceeding with the particular choice $b=\frac{1}{8}$ and solving the 
rest of the field equations, it turns out that the corresponding 
solution is surprisingly simple, so that $f(r)$ and $V(\phi)$ can
both be written in a neat and concise form. This simplified, critical solution 
will be the subject of concern in the rest part of this paper.

\section{Simplified solution and some of its properties}

In this and the subsequent sections, we shall consider only the case 
$b=\frac{1}{8}$. For convenience we shall also rewrite $k=\frac{1}{8B}$
and take $B$ as the only free parameter characterizing the scalar field. 
Clearly, we have 
\begin{align}
    \phi(r)=\pm \sqrt{\frac{8B}{r+B}}. \label{phisol}
\end{align}
Substituting this into the field equations, we get the following metric 
functions,
\begin{align}
    f(r)&=3\beta+\frac{2B\beta}{r}+\frac{(3r+2B)^2a^2}{r^4}+\frac{r^2}{\ell^2},
    \label{fsol}\\
    \omega(r)&=-\frac{(3r+2B)a}{r^3}, \label{omegasol}
\end{align}
where (\ref{omegasol}) is exactly the second term on the right hand side of
(\ref{omegab}). In order that $\phi(r)$ is nonsingular 
at finite $r$, it is necessary to require $B\ge 0$. The parameter $\beta$ is 
related to the black hole mass $M$ via
\begin{align}
  \beta=-\frac{M}{3}, \label{BM}
\end{align}
as can be seen in the $B=0$ limit, in which case the solution 
degenerates into the BTZ black hole. Without loss of generality, we shall assume 
$a\ge 0$ in the rest of the paper. The opposite choice $a<0$ can be recovered by 
a reflection of the angular coordinate, i.e. $\psi\to -\psi$.

Before carrying out explicit analysis on the structure of the spacetime given by 
the solution (\ref{phisol})-(\ref{omegasol}), let us first describe some of 
the geometric 
characteristics of the solution. First let us calculate some of the curvature 
invariants. The Ricci scalar is singular at $r=0$ if $a\ne 0$,
\begin{align*}
  R= -\left( \,{\frac {{30B}^{2}}{{r}^{6}}}+\,{\frac {36B}{{r}^{5}}}
 \right) {a}^{2}-\frac{6}{\ell^2}.
\end{align*}
Higher order curvature invariants, e.g. $R_{\mu\nu}R^{\mu\nu}$ and
$R_{\mu\nu\rho\sigma}R^{\mu\nu\rho\sigma}$ are also singular at $r=0$ even 
if $a=0$:
\begin{align*}
&R_{\mu\nu}R^{\mu\nu}= \frac{12}{{\ell}^{4}}
  +{\frac {6{B}^{2}{\beta}^{2}}{{r}^{6}}}+\left(
  {\frac {324{B}^{4}}{{r}^{12}}}+{\frac {792{B}^{3}}{{r}^{11}}} 
  +{\frac { 486{B}^{2}}{{r}^{10}}}\right) {a}^{4}\nonumber\\
  &\qquad\qquad + \left( {\frac {12{B}^{3}\beta}{{r}^{9}}}
  +{\frac {18{B}^{2}\beta}{{r}^{8}}}
  +{\frac {102{B}^{2}}{{r}^{6}{l}^{2}}}
  +{\frac {144B}{{r}^{5}{l}^{2}}}\right) {a}^{2},
  \\ 
&R_{\mu\nu\rho\sigma}R^{\mu\nu\rho\sigma}=  
  \frac{12}{{\ell}^{4}}+{\frac {24{B}^{2}{\beta}^{2}}{{r}^{6}}}
  +\left({\frac {396{B}^{4}}{{r}^{12}}}
  +{\frac {1008{B}^{3}}{{r}^{11}}} 
  +{\frac {648{B}^{2}}{{r}^{10}}}\right) {a}^{4}\nonumber\\
  &\qquad\qquad 
  + \left({\frac { 48{B}^{3}\beta}{{r}^{9}}}
  +{\frac {72{B}^{2}\beta}{{r}^{8}}}+{\frac {48{B}^{2}}{{r}^{6}{l}^{2}}}
  +{\frac {144B}{{r}^{5}{l}^{2}}}\right) {a}^{2}.
\end{align*}
Therefore, the metric contains a curvature singularity at $r=0$, which has to be 
enclosed by some event horizon in order for the solution to be physically 
acceptable. 

On the other hand, by direct evalation, we find that some 
of the components of the Cotton tensor
\[
C_{abc}=\nabla_c R_{ab}-\nabla_b R_{ac}+\frac{1}{4}\left(
\nabla_b R \, g_{ac}-\nabla_c R\, g_{ab}\right)
\]
are nonvanishing if $B > 0$. We give only one of the nonvanishing
components below:
\[
C_{\psi r \psi}=\left(\,{\frac {{12B}^{2}}{{r}^{5}}}+ \,{\frac {18B}{{r}^{4}}}
 \right) {a}^{2}+\,{\frac {3\beta\,B}{{r}^{2}}}.
\]
In $(2+1)$ dimensions, the nonvanishing Cotton tensor signifies that the metric 
is non conformally flat \cite{Garcia:2003bw}.

The scalar potential associated with the above solution is determined to be
\begin{align} 
  V(\phi)&= \frac{2}{\ell^2} + U(\phi),\nonumber\\
  U(\phi)&=X \phi^{6}+Y\,\frac {\left({\phi}^{6}-40\,{\phi}^{4}   
  +640\,{\phi}^{2}
  -4608\right){\phi}^{10}}{\left({\phi}^{2}-8\right)^{5}},   \label{pot}
\end{align} 
where $X,Y$ are given by\footnote{According to (\ref{XY}), it seems 
that the choice $B=0$ is not allowed, because in this case the scalar self 
coupling constants $X,Y$ diverges. However, we may substitute (\ref{phisol}) and 
(\ref{XY}) into (\ref{pot}) and then take the $B\to 0$ limit. Then we find that 
$U(\phi)$ actually vanishes as $B\to 0$, which is consistent with the fact that 
the scalar field $\phi$ is vanishing at $B=0$.}
\begin{align}
&X=\frac{1}{512}\left(\frac{1}{\ell^2}+\frac{\beta}{B^2}\right),\qquad 
Y=\frac{1}{512}\left(\frac{a^2}{B^4}\right). \label{XY}
\end{align}
This potential has a nice behavior in 
power series expansion,  
\begin{align}   U(\phi)\simeq X\,\phi^6 + Y\,O(\phi^{10}), 
\end{align} 
in which only even powers of $\phi$ are
present and all the coefficients in  the $O(\phi^{10})$ part are
positive as long as $Y>0$.  The $\phi^6$ potential in $(2+1)$-dimensional  
gravity was
previously known \cite{Nadalini:2007p2561}  to have good behavior in
yielding exact black hole  solutions. Our work shows that the
addition of the infinite higher even power series  will keep this
feature.

In order that the scalar potential to be bounded from below, we may have 
the following choices for the constants $X$ and $Y$:
\begin{itemize}
\item $Y=0$, $X\ge 0$;
\item $Y>0$, in which case $X$ can be an arbitrary real number. In 
particular,
\begin{itemize}
\item If $X\ge 0$, then $U(\phi)$ has only a single extremum, i.e. 
the minimum at $\phi=0$;
\item If $X<0$, then $U(\phi)$ has three extrema, i.e. one local maximum 
at $\phi=0$ with $U(0)=0$, and two minima at $\phi=\pm \phi_0$ for 
some $\phi_0$, with $U(\pm \phi_0)<0$. 
\end{itemize}
The qualitative behavior of the scalar potential is very similar to the
scalar potential found in our previous work \cite{Charged hairy}, though the 
concrete form of the potential is different. The similar qualitative
behavior of the scalar potential may be responsible for some universal 
properties of $(2+1)$-dimensional gravitational theories with a non-minimally 
coupled scalar field, such as the same CFT dual for theories with different 
potential functions.
\end{itemize}

\section{Horizon structure}

Now let us go back to the exact solution (\ref{phisol})-(\ref{omegasol}). If the 
curvature singularity at $r=0$ is not naked, $f(r)$ must contain some zeros 
which correspond to black hole horizons.  It can be seen from eq.(\ref{fsol}) that if 
$\beta \ge 0$, then $f(r)$ will be positive for all $r\in[0,+\infty)$. So, the 
existence of black hole horizons requires $\beta<0$. Moreover, eq.(\ref{fsol}) 
indicates that $f(r)\to + \infty$ as both $r\to 0$ and $r\to+\infty$. So, the 
function $f(r)$ must have at least one minimum, and the total number of extrema 
of $f(r)$ must be odd. Since the $a=0$ case of the solution is just the static 
hairy black hole solution which is studied before \cite{Charged hairy}, we will 
consider only the $a\ne 0$ solution below.

In order to identify the number of zeros of $f(r)$, we first need to know the 
exact number and values of its extrema.  The condition for the extrema of $f(r)$ 
is just the condition for the zeros of $f'(r)$. From the expression
\begin{align*}
  f'(r)=\,{\frac {2r}{{\ell}^{2}}}-\,{\frac {2\beta\,B}{{r}^{2}}}
  -\left( \frac{18}{r^3}+\,{\frac {36B}{{r}^{4}}}
  +\,{\frac {{16B}^{2}}{{r}^{5}}} \right) {a}^{2}
\end{align*}
we see that 
\begin{align*}
  f'(0)=-\infty, \quad f'(+\infty)=+\infty.
\end{align*}
So, $f'(r)$ has to have some zero in the range $r\in[0,+\infty)$.  
However, it is difficult to find the zeros of $f'(r)$ analytically. 
In order to gain some information about the number of zeros of $f'(r)$, 
let us first try to find the number of extrema of $f'(r)$ and the value of $f'(r)$
at its extrema. We denote by $r_i$ the location of possible extrema of $f'(r)$ 
(which need not exist at all). At the hypothetic extrema, we should have
\begin{align*}
&f'(r_i)= \,\frac {2r_{{i}}}{{\ell}^{2}}-\,{\frac {2\beta\,B}{{r_{{i}}}^{2}}}
  -\left( \frac{18}{{r_{{i}}}^{3}}+\,{\frac {36B}{{r_{{i}}}^{4}}}
  +\,{\frac{{16B}^{2}}{{r_{{i}}}^{5}}} \right) {a}^{2},\\
&f''(r_i)=\frac{2}{\,{\ell}^{2}}+\,{\frac {4\beta\,B}{{r_{{i}}}^{3}}}
  +\left(\,\frac{54}{r_{{i}}^4}+\,{\frac {144B}{{r_{{i}}}^{5}}}
  + \,{\frac {{80B}^{2}}{{r_{{i}}}^{6}}} \right) {a}^{2}=0.
\end{align*}
It is easy to see that
\begin{align*}
f'(r_i)&=\left(f'(r_i)+\frac{r_i}{2} f''(r_i)\right)
  = \,{\frac {3r_{{i}}}{{\ell}^{2}}}
  +\left( \frac{9}{\,{r_{{i}}}^{3}}+\,{\frac {36B}{{r_{{i}}}^{4}}}
  +\,{\frac {{24B}^{2}}{{r_{{i}}}^{5}}} \right) {a}^{2} >0.
\end{align*}
This means that if $f'(r)$ has some extrema, it must be positive at all 
the extrema. 
An alternative possibility is that $f'(r)$ has no extrema at all. In either 
cases the curve for $f'(r)$ will cross zero only once. Therefore  
$f(r)$ has only one minimum. Let us denote the location of the 
minimum of $f(r)$ as $r_{\mathrm{min}}$. Depending on the values of the 
parameters, we may encounter one of the following three possibilities: an 
extremal rotating hairy black hole, a non-extremal rotating hairy black hole or
a naked singularity.

\subsection{Extremal rotating hairy black hole}

The extremal rotating hairy black hole corresponds to the case in which $f(r)$ 
has only one zero at its minimum, i.e. $f(r_{\mathrm{min}})=0$ and 
$f'(r_{\mathrm{min}})=0$. The joint solution of these two equations is more 
restrictive than just the solution of $f'(r_{\mathrm{min}})=0$, so
we introduce a novel notation $r_{\mathrm{ex}}$ for the joint solution. If 
$r_{\mathrm{ex}}$ exists, then it will correspond to the horizon radius of the 
extremal rotating hairy black hole.

To actually get the horizon radius, we introduce
\begin{align}
P(r_{\mathrm{ex}}) &=\frac{B}{r_{\mathrm{ex}}+B}\, 
  \left( f(r_{\mathrm{ex}})+\frac{r_{\mathrm{ex}}(9r_{\mathrm{ex}}+6B)}{6B}
   f'(r_{\mathrm{ex}})\right) \nonumber\\
   &= -\left(\frac{27}{r_{\mathrm{ex}}^2}
   +\frac{36B}{r_{\mathrm{ex}}^3}
   +\frac{12B^2}{r_{\mathrm{ex}}^4}\right)a^2
   +\frac{3r_{\mathrm{ex}}^2}{\ell^2}=0, \label{p0}\\
K(r_{\mathrm{ex}}) &=\frac{r_{\mathrm{ex}}^4}{3B}P(r_{\mathrm{ex}})
   -\frac{r_{\mathrm{ex}}^5}{2B}f'(r_{\mathrm{ex}})
   =\beta r_{\mathrm{ex}}^3 + (6r_{\mathrm{ex}}+4B)a^2=0. \label{k0}
\end{align}
Then $r_{\mathrm{ex}}$ will be the joint solution of (\ref{p0}) and (\ref{k0}).

A necessary condition for (\ref{p0}) and (\ref{k0}) to have joint solution 
$r_{\mathrm{ex}}$ is
\begin{align}
  \beta&=-\frac{2a}{\ell}.\label{extremal condition}
\end{align}
Under this condition, 
$r_{\mathrm{ex}}$ is the real positive solution of the equation
\[
r_{\mathrm{ex}}^3-3a\ell\, r_{\mathrm{ex}}-2a\ell B=0,
\]
whose explicit value is given by
\begin{align}
r_{\mathrm{ex}}=z+\frac {a\ell}{z},\quad 
z=\left[a\ell \big(B
  +(B^2 -a\ell)^{1/2}\big)\right]^{1/3}  \label{extremal horizon}.
\end{align}

We can also think of eq.(\ref{p0}) as an equation for $B$. The solution reads
\[
B=-\frac{r_{\mathrm{ex}}}{2a\ell}\,\left(3a\ell \pm r_{\mathrm{ex}}^2\right).
\]
Since $B\ge 0$, only the minus sign choice in the bracket is allowed, and
we have
\[
r_{\mathrm{ex}}\ge \left(3a\ell\right)^{1/2},
\]
where the bound is saturated when $B=0$.

\subsection{Non-extremal rotating hairy black hole}

The non-extremal rotating hairy black hole corresponds to the conditions 
$f(r_{\mathrm{min}})<0$ and $f'(r_{\mathrm{min}})=0$. 
Therefore, $r_{\mathrm{min}}$ must obey the inequalities
\begin{align}
P(r_{\mathrm{min}}) <0, \quad K(r_{\mathrm{min}}) <0,
\end{align}
where $P(r)$ and $K(r)$ are given by the same expression in (\ref{p0})
and (\ref{k0}) but with $r_{\mathrm{ex}}$ changed into $r$. 
Since $P'(r)>0$ for all $r\in[0,+\infty)$, we see that $P(r)$ increases 
monotonically. So, the solution to the inequality $P(r)<0$ must 
be smaller than the solution of the equation 
$P(r)=0$, i.e $r_{\mathrm{min}}<r_{\mathrm{ex}}$.
Solving $\beta$ form the inequality $K(r_{\mathrm{min}}) <0$, we find that 
the necessary condition for the existence of two disjoint horizons is
\begin{align}
  \beta&<-\frac{2a}{\ell}.   \label{nonextremal condition}
\end{align}
However, due to the complicated form (\ref{fsol}) of the function $f(r)$, we are
unable to find its zeros $r=r_\pm$ analytically for generic choices of 
parameters. The only qualitative property we know 
about $r_\pm$ is that $r_- < r_{\mathrm{min}}, r_+>r_{\mathrm{min}}$, and that
$r_+$ is the event horizon for the non-extremal black hole.
However, for the specific choice $Y=-X>0$, we can indeed work out the explicit 
values of the two horizon radii. This will be shown in the next section.

\subsection{Naked singularity}

If $f(r)>0$ at its minimum $r_{\mathrm{min}}$, then there will be no horizons 
at all in the solution. In this case the curvature singularity at $r=0$ will be
naked. This can happen if
\begin{align}
  \beta&>-\frac{2a}{\ell},\label{naked singularity}
\end{align}
which can be obtained by analyzing the allowed values of $\beta$ as the solution
to the inequality $K(r)>0$. We will not expand our discussions in this case, 
because naked singularity is physically unappealing.
\vspace{8pt}

Summarizing the above three subsections, we find that the existence of black 
hole horizons imposes an upper bound for the parameter $\beta$, which reads
\[
\beta \le -\frac{2a}{\ell}.
\]
Since $\beta$ is related to the mass of the black hole, we can also view this as
a bound of the mass-angular momentum ratio,
\[
-\frac{\beta}{a}\ge \frac{2}{\ell}.
\]
Actually we can take this bound as an analogue of the famous Kerr bound in (2+1) 
dimensions.

\section{Special cases}

The scalar potential $U(\phi)$ contains two independent coupling constants $X$ 
and $Y$. If either (or both) of these constants vanish(es), tremendous 
simplification of the solution would occur. Moreover, taking special values for 
these constants or the parameters appearing the solution would also make the 
solution simplified, and there are cases in which the simplified versions of the 
solution have already been found in the literatures. In this section, we shall 
list some of the special cases of the solution and the corresponding conditions 
imposed on the parameters.

\subsection{Degenerate cases which are previously known}

In this subsection we list three degenerate cases which are known in the 
literatures. These are:
\begin{itemize}
\item Rotating BTZ black hole \cite{Banados:1992wn}, which corresponds to the
choice $B=0$. As mentioned in the footnote right beneath eq.(\ref{BM}), taking 
the $B=0$ limit effectively set both $X$ and $Y$ equal to zero, so, both the 
scalar field and its potential vanishes in this case. The solution reads
\begin{align*}
   f(r)&=3\beta+\frac{9a^2}{r^2}+\frac{r^2}{\ell^2},\\
   \omega(r)&=-\frac{3a}{r^2}.
\end{align*}

\item Static hairy black hole with a $\phi^6$ potential 
\cite{Henneaux:2002p2538,Nadalini:2007p2561}, which corresponds to the choice 
$a=0$ or $Y=0$. The solution reads
\begin{align*}
   f(r)&=3\beta+\frac{2B\beta}{r}+\frac{r^2}{\ell^2},\\
 \phi(r)&=\pm \left(\frac{8B}{r+B}\right)^{1/2},\\
  U(\phi)&=X {\phi}^{6},\\
   \omega(r)&=0.
\end{align*}

\item Conformally dressed black hole\cite{Martinez:1996gn}, which corresponds to 
$X=0$, $Y=0$ but $B\ne 0$. Notice that $Y=0$ implies $a=0$, so the solution 
becomes static. Explicitly,
\begin{align*}
   f(r)&=-\frac{3B^2}{\ell^2}-\frac{2B^3}{\ell^2}\,\frac{1}{ r}
   +\frac{r^2}{\ell^2},\\
   \phi(r)&=\pm \left(\frac{8B}{r+B}\right)^{1/2},
\end{align*}
and both $U(\phi)$ and $\omega(r)$ vanishes.
\end{itemize}
Since these degenerate cases are already known in the literatures, we will not 
make any discussion about their properties.

\subsection{Special cases which makes the solution simplify} 

In this subsection we shall look at a special case of the solution with
$Y=-X>0$, i.e. $\beta=-\frac{a^2\ell^2+B^4}{B^2\ell^2}$. We are interested in 
this particular choice of parameter because the metric function $f(r)$ possesses 
a particular factorized form, which enables us to calculate the horizon radii 
and other properties of the black hole solution in analytic form.  

The solution now reads
\begin{align}
   f(r)&=\frac{(r+B)^2 (r-2B)(B^2 r^3 - 3a^2 \ell^2 r -2B a^2 \ell^2)}
   {B^2 \ell^2 r^4}, \label{f}\\
   \omega(r)&=-\frac{(3r+2B)a}{r^3},\\
  \phi(r)&=\pm \left(\frac{8B}{r+B}\right)^{1/2}, \label{phisp}\\
  U(\phi)&=\frac{1}{512}\left(\frac{a^2}{B^4}\right)
  \left(- \phi^{6}+\frac {\left( {\phi}^{6}-40\,{\phi}^{4}
  +640\,{\phi}^{2}-4608 \right){\phi}^{10}}{\left( {\phi}^{2}-8\right) ^{5}}
  \right).
  \label{U}
\end{align}
The condition (\ref{extremal condition}) for extremal black hole becomes 
$B^2=a \ell$, with the horizon radius $r_{\mathrm{ex}}=2B$. 
More generally, for any $B> 0$, we have 
\[
-\beta-\frac{2a}{\ell}=\frac{(B^2-a\ell)^2}{B^2\ell^2}\ge 0,
\]
i.e. $\beta\le -\frac{2a}{\ell}$, in which the 
equality holds only in the extremal case. Therefore, 
under the choice $\beta=-\frac{a^2\ell^2+B^4}{B^2\ell^2}$, the solution will 
always behave as a black hole solution, and the singularity at the origin can 
never become naked.  The horizon radii can be worked out explicitly, which read
\begin{align*}
&r_1 = 2\left(\sigma a\ell\right)^{1/2},\quad 
r_2 = \left(\frac{a\ell}{\sigma}\right)^{1/2}\left(\zeta+\frac{1}{\zeta}\right),
\end{align*}
where we have set $B^2=\sigma a\ell$, and
\[
\zeta=\left(\sigma+(\sigma^2-1)^{1/2}\right) ^{1/3 }.
\]
If $0<\sigma<1$, $r_2$ is the outer horizon, if $\sigma>1$, $r_1$ is the outer 
horizon. The critical choice $a=1$ corresponds to the extremal case, with 
$r_1=r_2=r_{\mathrm{ex}}$.

Incidentally, the extrema of $U(\phi)$ can also be obtained analytically in this 
case. Solving the condition $\frac{dU}{d\phi}=0$ as an algebraic equation, we 
get either $\phi=0$, which corresponds to a five-fold degenerate local maxima, 
or $\phi=\pm\phi_0=\pm\frac{2\sqrt{3}}{3}$,  which correspond to a pair of 
reflection symmetric minima of $U(\phi)$. The value of $U(\phi)$ at the minima
reads
\[
U(\pm\phi_0)=-\frac{7}{3125}\left(\frac{a^2}{B^4}\right),
\]
which adds an extra negative contribution to the bare cosmological constant 
$-\frac{1}{\ell^2}$.

If we fix $\phi$ to take the constant value $\phi_0$ in the action, 
then the system becomes a pure Einstein-AdS gravity theory and the true 
cosmological constant will become 
\[
\Lambda_{\mathrm{eff}}=-\frac{1}{\ell^2}+U(\pm\phi_0)=-\frac{1}{\ell^2}
-\frac{7}{3125}\left(\frac{a^2}{B^4}\right) \equiv 
\frac{1}{\ell_{\mathrm{eff}}^2}.
\]
Notice that the effective cosmological constant $\Lambda_{\mathrm{eff}}$ is 
negative even if the bare value $-\frac{1}{\ell^2}$ vanishes, i.e. 
$\ell\to \infty$. The rotating BTZ black hole mentioned in the previous 
subsection is an exact solution to this theory, but with 
$\ell\to\ell_{\mathrm{eff}}$. However, this solution cannot 
be obtained from our solution (\ref{f})-(\ref{U}), because the scalar field 
$\phi$ as given in (\ref{phisp}) can never become a constant. 

What will happen if we take the $\ell\to\infty$ limit in the solution 
(\ref{f})-(\ref{U})? It is clear that $\phi(r), \omega(r)$ and $U(\phi)$ are not 
affected by this limit, but the outer horizon in the metric will be pushed to 
$r=+\infty$, because the $r^3$ term in the last factor in $f(r)$ drops off while 
taking the limit. Consequently, there will be no ``outside region'' of the black 
hole. Any observer must be located either inside the inner 
horizon with $r<r_1=2B$, or in between the two horizons with $2B<r<+\infty$.
In the former case, the spacetime patch in which the observer is inhabited is 
stationary, with a ``cosmological horizon'' at $r=2B$, but 
also contains a naked singularity at the origin, which is physically 
uninteresting. In the latter case, the function $f(r)$ becomes negative in the 
spacetime patch the observer is located in, so, the coordinate $r$ becomes 
timelike and $t$ spacelike. The metric is time dependent in this patch, with two 
temporal horizons at $r=2B$ and 
$r=+\infty$ respectively. This means that time has both a beginning and an end. 
Moreover, the metric is no longer rotating, but with some spacial twisting, 
because the non-diagonal elements of the metric live purely in the spacial 
directions. The spacetime in this situation cannot be understand as a usual 
black hole solution.

\section{Concluding remarks}

The study of $(2+1)$-dimensional gravity with matter source turns out to be very 
fruitful and many exact solutions with black hole solutions have been found in 
the recent years. Naturally one expects that such solutions will play some role 
in the forthcoming constructions for the dual theories in the spirit of AdS/CFT 
duality. The present work add some more input in this picture.

It is worth mentioning that although the solution we presented in this paper is 
a rotating hairy black hole without electromagnetic charge, the scalar field
takes exactly the same value as in the solution to the static charged hairy 
black hole case. This coincidence of scalar solution actually lies in the heart 
of the whole construction. As long as we impose the condition that $\phi$ 
depends only on $r$, the sourced Einstein's equation will imply that the value 
of the scalar field $\phi$ is forced to take the form (\ref{scsol}), of which 
(\ref{phisol}) is a particular choice. Moreover, this procedure actually 
determines the allowed type of the scalar potential $U(\phi)$
up to several adjustable constants. 
This phenomenon has also been observed in other studies of 
hairy black holes in four- \cite{Anabalon:2009p2560,Anabalon:2012p2557,
Anabalon:2012p2562} and five-dimensions \cite{Astefanesei}. 
The lesson to be learnt here is that, 
interaction between gravity and the scalar 
field imposes strong constraints on the allowed form of scalar potential. Such 
constraints, when applied in more realistic field theoretic models, will greatly 
help us in understanding the dynamics of fundamental scalar matter 
which is otherwise difficult to determine in flat spacetime.

After the first version of this paper has appeared in arXiv, we were notified of 
the reference \cite{Aparicio}, in which an algorithm for generating stationary 
axi-symmetric solutions of 3D dilaton gravity in the Einstein frame was 
proposed. In there, the scalar potential in the Einstein frame with the
ADM metric ansatz can contain a genenral function besides several integration 
functions. It will be interesting to see the relationship between the result
of \cite{Aparicio} with ours.

\providecommand{\href}[2]{#2}\begingroup
\footnotesize\itemsep=0pt
\providecommand{\eprint}[2][]{\href{http://arxiv.org/abs/#2}{arXiv:#2}}
\endgroup

\end{document}